\documentclass{PoS}
\usepackage{amssymb}
\usepackage{cite}
\usepackage{graphicx}
\usepackage{url}
\usepackage{lineno}

\title{Anisotropic Flow Measurements of Identified Particles in the STAR Experiment}

\ShortTitle{Anisotropic Flow Measurements of Identified Particles in the STAR Experiment}

\author{\speaker{Shaowei Lan$^{1,2}$} (for the STAR Collaboration)\\
        $^1$Central China Normal University\\
        $^2$Lawrence Berkeley National Laboratory\\
        E-mail: \email{shaoweilan@mails.ccnu.edu.cn}}

\abstract{
We report results of $v_1$ and $v_2$ for identified hadrons ($\pi^{\pm}, K^{\pm}, K_{S}^{0}, p, \phi$ and $\Lambda$) from Au+Au collisions at $\sqrt{s_{NN}}$ = 3 GeV and $v_2$ for $\pi^{\pm}, K^{\pm}, p$ and $\bar{p}$ at $\sqrt{s_{NN}}$ = 27 and 54.4 GeV using the STAR detector at RHIC. In Au+Au collisions at high energies, one finds that the values of $v_2$ are all positive and the number-of-constituent-quark (NCQ) scaling holds. On the other hand, results from collisions at 3 GeV, the midrapidity $v_2$ is negative for all hadrons and NCQ scaling is absent. In addition, the midrapidity $v_1$ slopes for all hadrons are found to be positive. Furthermore, the features of negative $v_2$ and positive $v_1$ slope at 3 GeV can be reproduced by calculations with baryonic mean-field potential in transport model. These results imply that in 3 GeV Au+Au collisions, the medium is characterized by baryonic interactions.

}

\FullConference{Critical Point and Onset of Deconfinement - CPOD2021\\
		15-19 March, 2021\\}
\begin{document}


\section{Introduction}
One of the most important motivations of relativistic heavy-ion collisions is to explore the Quantum Chromodynamics (QCD) phase diagram~\cite{Adams:2005dq}.
Lattice QCD calculations show that the phase transition from hadronic matter into the Quark-Gluon Plasma (QGP) phase is a smooth crossover at vanishing baryon chemical potential region, while a first-order phase transition is expected at the finite baryon chemical potential region~\cite{Aoki:2006we}.
Searching for the onset of QGP and studying the properties of QCD medium is the focus of the ongoing RHIC Beam Energy Scan Phase II (BES-II) program. 

Anisotropic flows~\cite{PhysRevLett.78.2309,PhysRevD.46.229}, the first two components of which are called directed flow ($v_1$) and elliptic flow ($v_2$), are effective tools to study the properties of the QCD matter created in heavy-ion collisions. They are generated early in the system evolution and are sensitive probes of the Equation-of-State (EoS) of the produced medium. Based on the hydrodynamical calculations, along the collision energy ($\sqrt{s_{NN}}$) the minimum in midrapidity $v_1$ slopes is proposed as a signal of the first-order phase transition between hadronic matter and the QGP~\cite{Hung:1994eq,Stoecker:2004qu}. In the Beam Energy Scan Phase I (BES-I) program, STAR reported that the $v_1$ slopes for identified particles as a function of collision energy~\cite{PhysRevLett.112.162301,PhysRevLett.120.062301}. A nonmonotonic behavior of midrapdiity $v_1$ slopes as a function of collision energy for protons and $\Lambda$ is observed and the $v_1$ slopes reach a minimum around $\sqrt{s_{NN}}$ = 10-20 GeV. On the other hand, large positive $v_2$ along with the observation of its number-of-constituent-quark (NCQ) scaling are strong evidence for the formation of a hydrodynamically expanding QGP phase with partonic degree of freedom~\cite{PhysRevLett.116.062301,PhysRevLett.118.212301}. Positive $v_2$ of hadrons at midrapidity has been observed from the top RHIC energy down to 4.5 GeV~\cite{PhysRevC.93.014907,PhysRevC.103.034908}. The $v_2$ values are found to be negative at $\sqrt{s_{NN}}$ $\leq$ 3.6 GeV due to the shadowing effect by the passing spectator nucleons~\cite{PhysRevLett.83.1295}.  Previous studies have shown that the $v_1$ and $v_2$ are particularly sensitive to nuclear incompressibility ($\kappa$) in the high baryon density region~\cite{Danielewicz:2002pu}. The constrains on $\kappa$ by comparing experimental data with results from the theoretical transport model will certainly help us to understand nuclear EoS.

In this paper, we present new results of $v_1$ and $v_2$ for identified hadrons ($\pi^{\pm}, K^{\pm}, K_{S}^{0}, p, \phi$ and $\Lambda$) at $\sqrt{s_{NN}}$ = 3 GeV and $v_2$ of $\pi^{\pm}, K^{\pm}, p$ and $\bar{p}$ at $\sqrt{s_{NN}}$ = 27 and 54.4 GeV from the STAR experiment at RHIC. 

\section{Analysis Details}
Minimum-bias triggered data with 260M, 560M and 600M events from $\sqrt{s_{NN}}$ = 3, 27 and 54.4 GeV Au+Au collisions, respectively, are used in this analysis.
The Time Projection Chamber (TPC)~\cite{tpc:anderson} is the main detector of STAR which performs charged particle tracking near midrapidity. 
The 3 GeV data were taken, with beam energy of 3.85 GeV per nucleon, in 2018 in the fixed-target (FXT) mode. The target is positioned inside the beam pipe near the edge of the TPC, at 200 cm from the TPC center along the beam axis. This gives an experimental acceptance coverage of 0 $\textless$ $\eta$ $\textless$ 2 in pseudorapidity. The higher energy data were taken in the collider mode, where the beam bunch crossing restricted to the TPC central region, yielding an acceptance of $|\eta|$ $\textless$ 1. Particle identification is based on the energy loss information from TPC detector and time of flight information from Time-of-Flight (TOF) detector~\cite{tof:geurts}. At $\sqrt{s_{NN}}$ = 3 GeV, the first-order event plane is determined with the Event Plane Detector (EPD)~\cite{epd:adams} located on the east side of the STAR detector system. The $v_1$ and $v_2$ are calculated with the first-order event plane. For $\sqrt{s_{NN}}$ = 27 and 54.4 GeV, the second-order event plane is reconstructed from tracks recorded by the TPC. In order to avoid self-correlation and suppress non-flow effects, a $\eta$-subevent plane method is used for the $v_2$ calculation, in which an $\eta$ gap of $|\Delta \eta|$ $=$ 0.1 between positive and negative $\eta$-subevent is applied~\cite{epmethod}.

\section{Results and Discussion}

\begin{figure}[!htb]
    \centering
    \centerline{\includegraphics[width=0.7\textwidth]{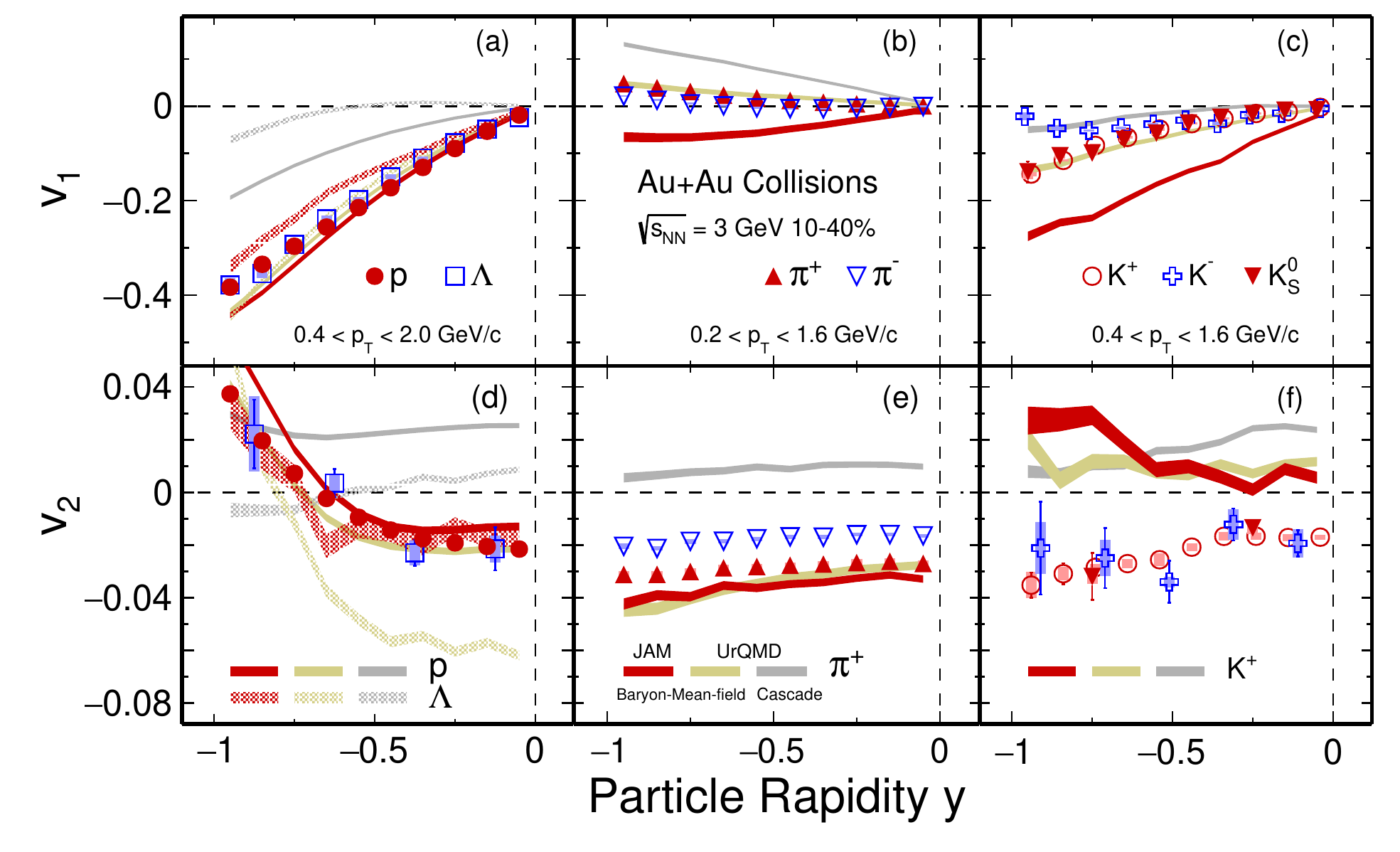}}
    \caption{(Color online) Rapidity dependence of $v_1$ (top panel) and $v_2$ (bottom panel) for $p, \Lambda, \pi^{\pm}, K^{\pm}$ and $K^{0}_{S}$ in 10-40\% Au+Au collisions at $\sqrt{s_{NN}}$ = 3 GeV. }
    \label{fig:vny_3gev}
\end{figure}

The rapidity dependence of directed flow ($v_{1}$) and elliptic flow ($v_{2}$) for identified hadrons ($p, \Lambda, \pi^{\pm}, K^{\pm}$ and $K^{0}_{S}$) from the $\sqrt{s_{NN}}$ = 3 GeV 10-40\% centrality is shown in top and bottom panels in Fig.~\ref{fig:vny_3gev}, respectively. Due to the acceptance, the results from the rapidity region $-$1 $\textless$ $y$ $\textless$ 0 are presented and the corresponding $p_T$ range for each hadron is shown in the figure. From the top panel, the baryons' $v_{1}$ in panel (a) is significantly larger in magnitude than mesons'. Small charge dependence of $v_1$ has been observed for pions in panel (b) and kaons in panel (c). As shown in the lower panel, the measured $v_2$ for all hadrons at midrapidity ($|y| \leq$0.5) are negative, implying an out-of-plane expansion in the 3\,GeV collision, contrary to the in-plane expansion in high energy collisions. 
For comparison, the calculations from transport model UrQMD~\cite{urqmd} and JAM~\cite{jamref}, with same centrality and kinematic selections for given hadrons, are shown as colored bands. Red, golden and gray bands represent the results of JAM mean-field, UrQMD mean-field and cascade mode, respectively. As one can see in the figure, the UrQMD and JAM calculations with baryonic mean-field potential can qualitatively describe the data, but not for $K^{+}$ results. The incompressibility value ($\kappa$) used in the mean-field option is 380 MeV in this analysis.

\begin{figure}[!htb]
    \centering
    \centerline{\includegraphics[width=0.7\textwidth]{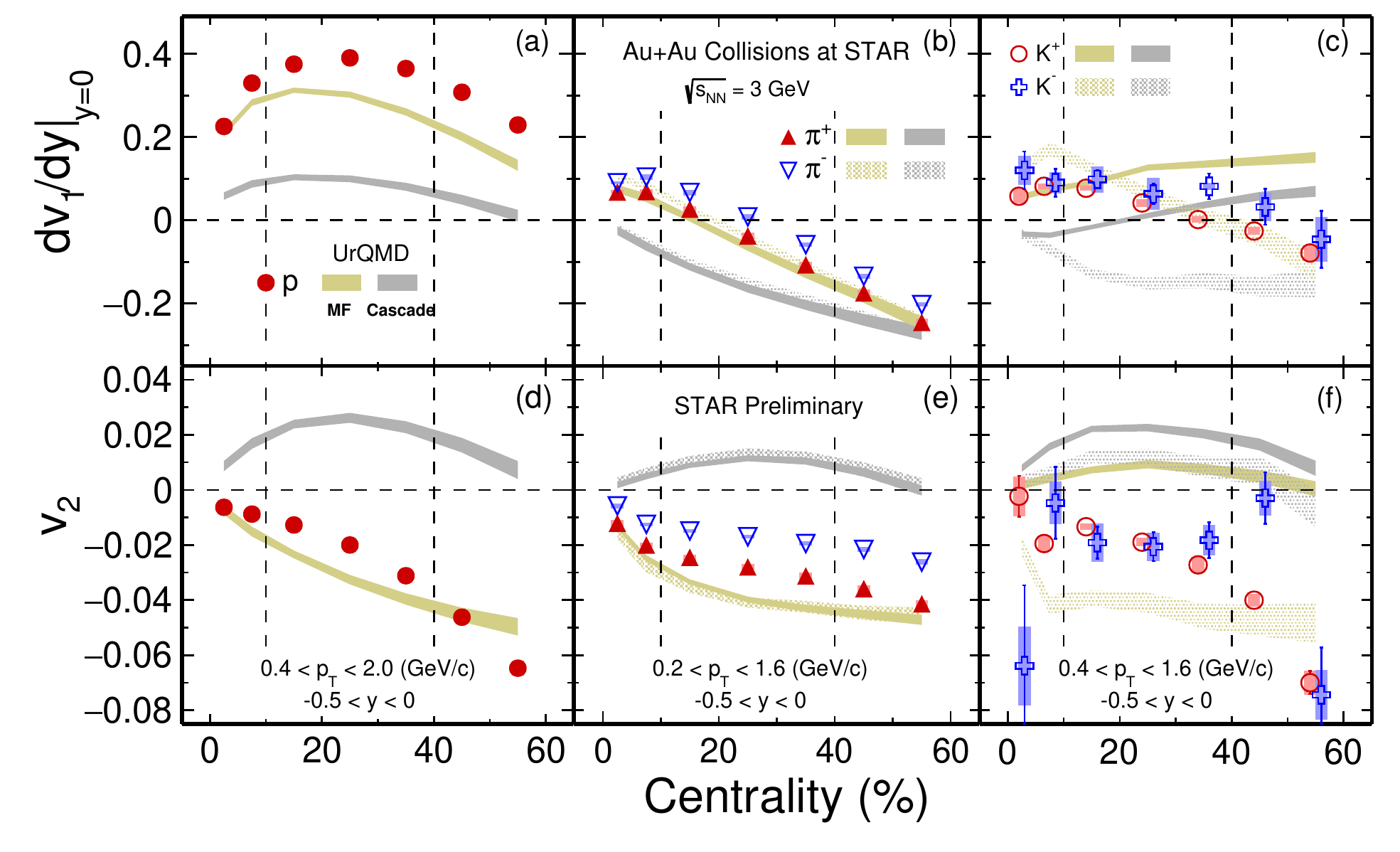}}
    \caption{(Color online) Centrality dependence of $dv_{1}/dy|_{y=0}$ (top panel) and $v_2$ (bottom panel) for $p, \pi^{\pm}$ and $K^{\pm}$ in Au+Au collisions at $\sqrt{s_{NN}}$ = 3 GeV. }
    \label{fig:vncent_3gev}
\end{figure}

We employed a polynomial fit of the function $v_{1} = a + by + cy^{3}$ to extract the strength of directed flow at midrapidity.
We refer to $dv_{1}/dy|_{y=0}$ as the slope obtained from the above fits. Figure~\ref{fig:vncent_3gev} shows the centrality dependence of $dv_{1}/dy|_{y=0}$ and $v_2$ at midrapidity for $p, \pi^{\pm}$ and $K^{\pm}$ in Au+Au collisions at $\sqrt{s_{NN}}$ = 3 GeV. The centrality dependence for $dv_{1}/dy|_{y=0}$ and $v_2$ of all hadrons has been observed.
The $v_2$ values at midrapidity for all hadrons are found negative.
The results from UrQMD mean-field and cascade mode are represented by the golden and gray bands, respectively.  The UrQMD results with baryonic mean-field potential can qualitatively describe the data, but not for $K^{+}$ results.

\begin{figure}[!htb]
    \centering
    \centerline{\includegraphics[width=0.7\textwidth]{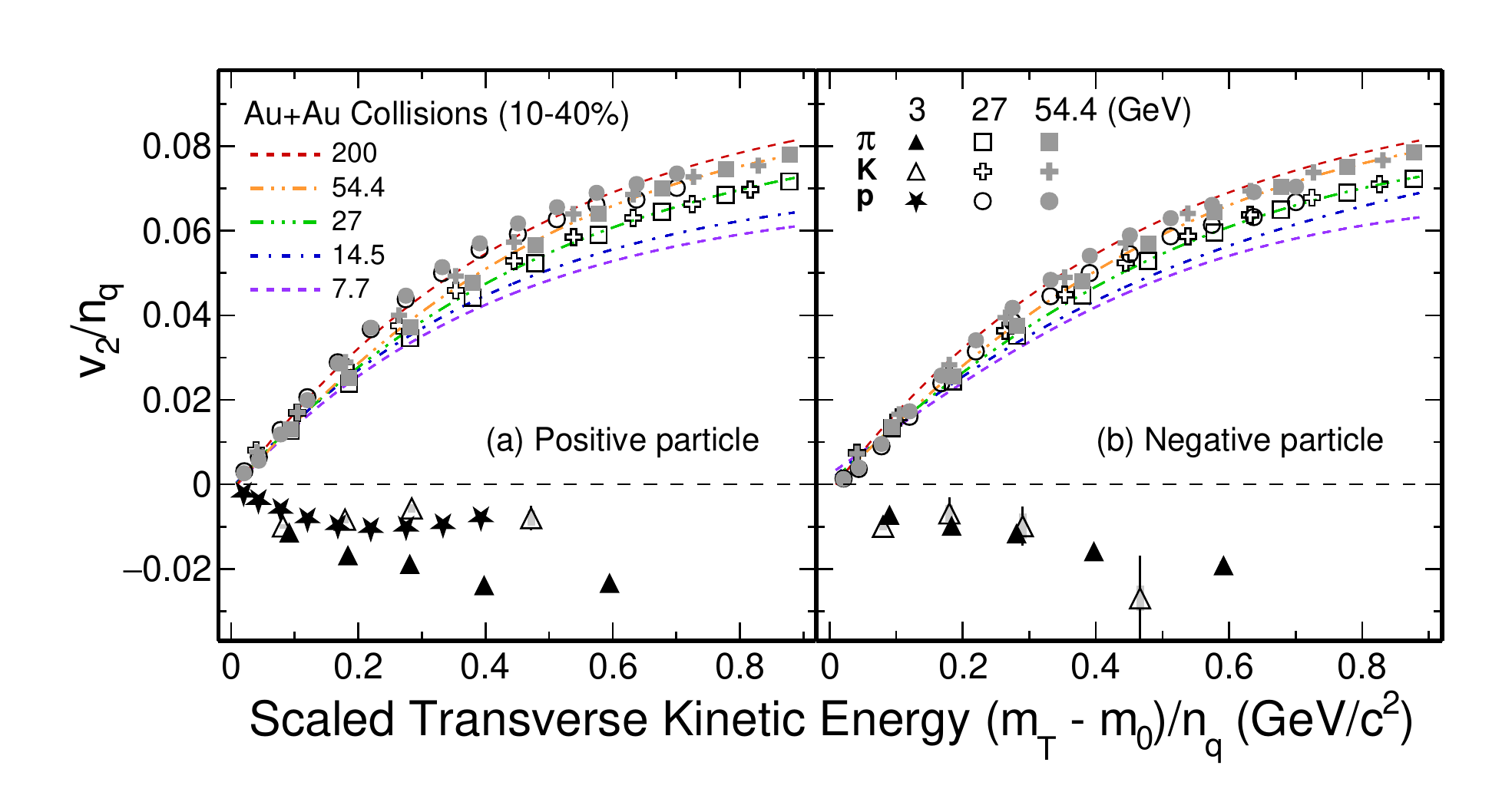}}
    \caption{(Color online) NCQ scaled $v_2$ as a function of $(m_{T}-m_{0})$/$n_{q}$ for $\pi^{\pm}, K^{\pm}, p$ and $\bar{p}$ from Au+Au collisions in 10-40\% centrality at $\sqrt{s_{NN}}$ = 3, 27 and 54.4\,GeV for positive (left panel) and negative (right panel) charged particles.}
    \label{fig:v2ncq_energy}
\end{figure}

Figure~\ref{fig:v2ncq_energy} presents $v_2$ scaled by the number of constituent quarks, $v_{2}/n_{q}$, as a function of scaled transverse kinetic energy, $(m_{T}-m_{0})/n_{q}$, for $\pi^{\pm}$ and $K^{\pm}, p$ and $\bar{p}$ in 10-40\% midcentral Au+Au collisions at 3, 27 and 54.4 GeV. At the two higher energies, the NCQ scaling of $v_2$ holds well, which indicates that the collectivity is developed during the partonic phase stage. The colored dashed lines represent the scaling fit to data for both positive and negative charged particles at 7.7, 14.5, 27, 54.4 and 200 GeV~\cite{fitv2}. While for 3 GeV collisions, it is apparent that all values of $v_{2}/n_{q}$ are negative. Contrary to the higher energy data shown, the NCQ scaling disappears in such low energy collisions. As shown in Fig.~\ref{fig:vny_3gev} and Fig.~\ref{fig:vncent_3gev}, the JAM and UrQMD model calculations with baryonic mean-field potential reproduce the observed negative values of $v_2$ for baryons as well as pions. The new results clearly indicate different properties for the matter produced in the 3 GeV collisions. In other words, partonic interactions no longer dominate and baryonic scatterings take over.

\begin{figure}[!htb]
    \centering
    \centerline{\includegraphics[width=0.45\textwidth]{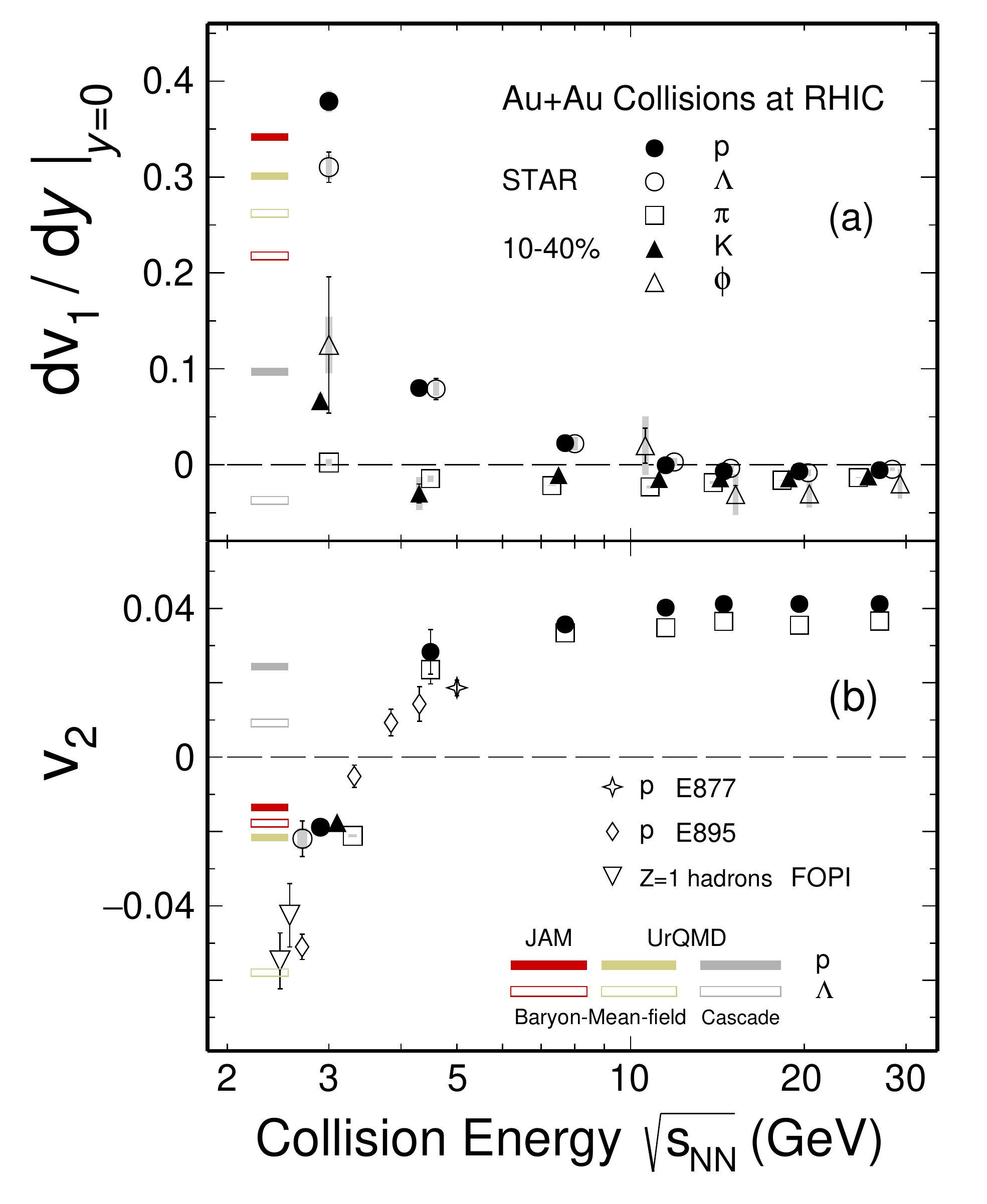}}
    \caption{(Color online) $v_1$ slopes $dv_{1}/dy|_{y=0}$ (top panel) and $v_2$ (bottom panel) versus $\sqrt{s_{NN}}$ for identified particles in heavy-ion collisions.}
    \label{fig:vnenergy}
\end{figure}

The collision energy dependence of $v_1$ slopes ($dv_{1}/dy|_{y=0}$) and $v_{2}$ for identified particles at 10-40\% centrality is summarized in Fig.~\ref{fig:vnenergy}. The top panel  shows $p_{T}$ integrated midrapidity $v_1$ slopes as a function of collision energy for $\pi, K, p, \phi$ and $\Lambda$ from Au+Au collisions at 10-40\% centrality interval. Here the positive and negative particle results are combined for both pions and kaons, as the difference between them is small, as shown in Fig.~\ref{fig:vny_3gev}. The new measurements from STAR at 3\,GeV for $\pi, K, p, \phi$ and $\Lambda$ are shown in the figure. The midrapidity $v_1$ slopes and $v_2$ for all hadrons are found to be positive and negative at 3 GeV, respectively.
The bottom panel in Fig.~\ref{fig:vnenergy} shows the midrapidity $v_2$ results as a function of collision energy for $\pi, p$ and charged hadrons from STAR, E877, E895 and FOPI~\cite{PhysRevC.103.034908,PhysRevC.93.014907,PhysRevLett.83.1295,fopiv2}.
Results from JAM and UrQMD calculations, with the same centrality and kinematic cuts used in data analysis, are shown as colored bands. Red, golden and gray bands represent the results of JAM mean-field, UrQMD mean-field and cascade mode, respectively. By including the baryonic mean-field potential, the JAM and UrQMD models reproduce the trends for both $dv_{1}/dy|_{y=0}$ and $v_2$ for baryons. It indicates that the dominant degrees of freedom are the interacting baryons in the 3 GeV collisions.

\section{Summary}

In summary, we reported the measurements of $v_1$ and $v_2$ for identified hadrons ($\pi^{\pm}, K^{\pm}, K_{S}^{0}, p, \phi$ and $\Lambda$) from Au+Au collisions at $\sqrt{s_{NN}}$ = 3 GeV and $v_2$ for $\pi^{\pm}, K^{\pm}, p$ and $\bar{p}$ at $\sqrt{s_{NN}}$ = 27 and 54.4 GeV from STAR experiment. 
The NCQ scaling of $v_2$ is observed for the collision energies $\geq$ 7.7 GeV. Due to the formation of the QGP, one finds that each hadron's $v_2$ is positive and all slopes of $v_1$ are negative. For the 3 GeV collisions, the NCQ scaling is absent and the opposite collective behavior is observed, namely, in contrast to that from high energy collisions, the values of $v_2$ for all hadrons are negative while the $v_1$ slopes are all positive. Transport model JAM and UrQMD calculations with baryonic mean-field potential qualitatively reproduce these results. These observations imply the vanishing of partonic collectivity and a new EoS, dominated by baryonic interactions in the high baryon density region.

\section*{Acknowledgements}
This work is supported in part by the National Key Research and Development Program of China under Grant No. 2020YFE0202002, the National Natural Science Foundation of China (under Grant No. 11890710, 11890711 and 12175084) and the Chinese Scholarship Council (CSC).

\end{document}